\documentclass[12pt,english]{article}
\usepackage[T1]{fontenc}
\usepackage{geometry}
\usepackage{color}
\usepackage{babel}
\usepackage{array}
\usepackage{mathtools}
\usepackage{multirow}
\usepackage{amsmath}
\usepackage{amsthm}
\usepackage{amssymb}
\usepackage{rotating}
\usepackage{rotfloat}
\usepackage{setspace}
\usepackage{appendix}

\usepackage[authoryear]{natbib}
\usepackage{CJKutf8}
\usepackage[unicode=true,pdfusetitle,
 bookmarks=true,bookmarksnumbered=false,bookmarksopen=false,
 breaklinks=false,pdfborder={0 0 0},pdfborderstyle={},backref=false,colorlinks=true]
 {hyperref}
\hypersetup{citecolor=blue}

\makeatletter

\theoremstyle{remark}

\theoremstyle{plain}

\theoremstyle{plain}

\theoremstyle{remark}

\theoremstyle{plain}

\usepackage{geometry}
\usepackage{fancyhdr}
\usepackage{bm}
\usepackage{alltt}

\usepackage[T1]{fontenc}
\IfFileExists{upquote.sty}{\usepackage{upquote}}{}
\usepackage[sc]{mathpazo}
\usepackage{indentfirst}
\usepackage{amsfonts}
\allowdisplaybreaks
\usepackage{authblk}

\makeatother

\providecommand{\assumptionname}{Assumption}
\providecommand{\notationname}{Notation}
\providecommand{\propositionname}{Proposition}
\providecommand{\remarkname}{Remark}
\providecommand{\theoremname}{Theorem}

\usepackage{float}
\usepackage{graphicx}
\usepackage{subfigure}
\usepackage{fancyhdr}
\usepackage{booktabs}
\usepackage{threeparttable}
\usepackage{array}
\usepackage{tabularx}
\usepackage{multirow}
\usepackage{ragged2e}
\usepackage{caption} 
\usepackage[labelsep=period]{caption}
\usepackage{subcaption}
\usepackage{makecell}
\usepackage{geometry}
\usepackage{amsmath}
\newcolumntype{L}{>{\raggedright\arraybackslash}X} 
\newcolumntype{C}{>{\centering\arraybackslash}X} 
\usepackage{float}
\usepackage{afterpage}
\usepackage{hyperref}

\vspace{-1em}
\title{\textbf{Artificial Intelligence and Skills: Evidence from Contrastive Learning in Online Job Vacancies}%
\thanks{Hangyu Chen: Zhejiang University, \href{mailto:hangyuchen@zju.edu.cn}{\tt \textcolor{blue}{hangyuchen@zju.edu.cn}}; Yongming Sun: Zhejiang University, \href{mailto:yongming.sun@zju.edu.cn}{\tt \textcolor{blue}{yongmingsun@zju.edu.cn}}; Yiming Yuan (corresponding author): Zhejiang University, \href{mailto:yuanyiming@zju.edu.cn}{\tt \textcolor{blue}{yuanyiming@zju.edu.cn}}. We acknowledge financial support from the National Social Science Fund of China (72141305). All authors contributed equally to this work, and authors are listed in alphabetical order by last name. All errors are our own.}}

\author[1,2]{Hangyu Chen}
\author[3]{Yongming Sun}
\author[3]{Yiming Yuan}

\affil[1]{Research Center for Regional Coordinated Development, Zhejiang University, China}
\affil[2]{Center for Research on China Open Economy, Zhejiang University, China}
\affil[3]{School of Economics, Zhejiang University, China}

\date{January 2026}
\vspace{-2em}
\begin{document}

\maketitle

\begin{abstract}
\vspace{-1em}
\noindent We investigate the impact of artificial intelligence (AI) adoption on skill requirements using 14 million online job vacancies from Chinese listed firms (2018–2022). Employing a novel Extreme Multi-Label Classification (XMLC) algorithm trained via contrastive learning and LLM-driven data augmentation, we map vacancy requirements to the ESCO framework. By benchmarking occupation-skill relationships against 2018 O*NET-ESCO mappings, we document a robust causal relationship between AI adoption and the expansion of skill portfolios. Our analysis identifies two distinct mechanisms. First, AI reduces information asymmetry in the labor market, enabling firms to specify current occupation-specific requirements with greater precision. Second, AI empowers firms to anticipate evolving labor market dynamics. We find that AI adoption significantly increases the demand for "forward-looking" skills—those absent from 2018 standards but subsequently codified in 2022 updates. This suggests that AI allows firms to lead, rather than follow, the formal evolution of occupational standards. Our findings highlight AI’s dual role as both a stabilizer of current recruitment information and a catalyst for proactive adaptation to future skill shifts.\\


\end{abstract}

\thispagestyle{empty}

\newpage
\setcounter{page}{1}

\section{Introduction} \label{sec:Introduction}

\noindent The rapid diffusion of artificial intelligence (AI) is transforming labor markets in ways that extend well beyond the automation of tasks. A growing body of research documents AI's effects on employment, wages, and occupational structure, primarily through the lens of task displacement and skill-biased technological change \citep{AcemogluRestrepo2018,AcemogluRestrepo2020, AutorLevy2003, FreyOsborne2017}. Yet labor markets are characterized not only by production relationships but also by pervasive search and matching frictions. Workers and firms must find each other, communicate their attributes, and form matches under imperfect information \cite{RogersonShimer2005}. How AI affects these information flows—and the signaling mechanisms through which market participants reveal their characteristics—remains far less understood.

Signaling is fundamental to labor market functioning. On the worker side, job seekers invest in education, craft application materials, and accumulate credentials to communicate their abilities to prospective employers \citep{Spence1973, AryalBhuller2022}. On the employer side, firms signal their requirements through job postings that specify occupational titles, responsibilities, and skill demands. These signals shape the pool of applicants, influence match quality, and ultimately determine labor market efficiency. When signals are precise and informative, workers can better target suitable positions, and firms can attract candidates whose capabilities align with job requirements. When signals are noisy or incomplete, both sides incur wasteful search costs and matches deteriorate.

Recent evidence suggests that generative AI may be disrupting signaling on the job seeker side of the market. \citet{GaldinSilbert2025} show that large language models have substantially lowered the cost of producing written application materials, eroding the informativeness of cover letters as signals of worker quality. In their analysis, AI-assisted applications become more generic and less predictive of ability, causing labor market outcomes to become less meritocratic. Similarly, \citet{CuiDias2025} find that while AI cover letter tools increase textual alignment with job posts and raise callback rates, they simultaneously reduce the correlation between application quality and hiring success—consistent with signal dilution. \citet{WilesHorton2025} document parallel effects on the employer side of search costs: when firms receive AI-generated drafts of job postings, they post more jobs but produce less informative descriptions, ultimately harming match formation and generating welfare losses.

These findings raise a natural question: if AI can degrade signaling quality when it merely reduces the cost of producing signals, can AI also enhance signaling quality when it improves firms' underlying understanding of what they need? The distinction is crucial. The existing evidence focuses on settings where AI assists with communication—helping agents write better-sounding text—without necessarily improving their knowledge of the relevant attributes to communicate. But AI may also function as an information technology that helps firms process complex internal data, understand evolving job requirements, and articulate skill demands with greater precision.

This possibility is particularly relevant for employer signaling through job vacancies. Firms face a nontrivial challenge when designing job postings: they must translate internal organizational knowledge about tasks, workflows, and competencies into explicit skill requirements that prospective workers can understand and respond to. This translation is imperfect. Job content evolves continuously as technologies change and tasks are reorganized, yet occupational titles and standard classification systems update only periodically \cite{Autor2001}. Firms may lack the capacity to fully articulate the skills their positions require, especially for roles undergoing rapid transformation. If AI capability improves firms' ability to process job-related information and specify requirements, it could reduce information asymmetry between employers and job seekers—enhancing rather than degrading the quality of labor market signals.

The matching literature has long emphasized that skill requirements are endogenous to labor market conditions. \citet{AlbrechtVroman2002} model how firms choose skill requirements when jobs differ in complexity and workers differ in ability, showing that the equilibrium mix of job types depends on the costs of search and the distribution of worker skills. \citet{ModestinoShoag2016} provide empirical evidence that firms adjust skill requirements in response to labor supply: during the Great Recession, when workers were plentiful, firms raised education and experience requirements—a phenomenon they term "upskilling." These findings establish that skill requirements are strategic choices that respond to market conditions and information. Yet we know little about how technological capabilities within firms shape these choices.

A related literature examines how technological change affects the skill content of jobs. \citet{BraxtonTaska2023} use detailed skill requirements from online vacancies to study earnings losses following job displacement, finding that technological change accounts for nearly half of post-displacement earnings declines by requiring workers to acquire new skills. \citet{DemingKahn2018} document the rise of social skills in job postings and show that skill requirements predict wage levels and firm performance. These studies demonstrate the value of analyzing skill requirements at granular levels, but they focus on aggregate trends rather than firm-level responses to technology adoption.

In this paper, we investigate how firm-level AI capability affects the skill content of job postings—a direct measure of employer signaling in the labor market. We focus on China, a rapidly digitizing economy where both AI adoption and online recruitment have expanded at exceptional scale, providing an ideal setting to study these dynamics. Our analysis draws on three primary data sources: over 14 million online job postings from major recruitment platforms (Zhaopin, 51job, and Liepin) covering 2018–2022; AI patent records from the Incopat global patent database to measure firm-level AI capability; and firm financial data from the China Stock Market and Accounting Research (CSMAR) database to construct controls. By matching these datasets to Chinese listed firms and their subsidiaries, we construct a firm–occupation–year panel that links AI adoption to granular measures of skill demand.

A central empirical challenge is measuring skill requirements in a consistent and comparable manner across millions of job postings. Job vacancy texts are unstructured, vary widely in format and verbosity, and often express skill requirements implicitly rather than explicitly. For instance, a posting stating "ability to independently publish in high-impact international journals" implies demands for English proficiency and academic writing skills without naming them directly \cite{Bhola2020}. Traditional approaches based on keyword matching or named entity recognition (NER) fail to capture such implicit requirements and produce skill counts that are not comparable across postings with different writing styles. A posting requiring "proficiency in Python and R for data analysis" would be tagged with three skills, while one requiring "proficiency in relevant software for data analysis" would receive only one—despite the latter plausibly demanding at least the same skill set.

To address these challenges, we develop a novel skill extraction framework based on Extreme Multi-Label Classification (XMLC). Following recent advances in the NLP literature \citep{Liu2017, Decorte2023}, we frame skill identification as a classification problem where each job posting sentence can be mapped to multiple labels from a pre-defined skill taxonomy. We adopt the European Skills, Competences, Qualifications and Occupations (ESCO) framework, which provides a comprehensive and standardized enumeration of 13,939 distinct skills. Our approach employs a bi-encoder architecture trained via contrastive learning, where job posting sentences and ESCO skill labels are embedded into a shared semantic space. To overcome the absence of large-scale labeled training data, we leverage large language models (LLMs) to generate synthetic training pairs—skill labels matched with realistic job advertisement sentences that would imply those skills. This allows the model to learn semantic associations between skill concepts and their natural language expressions, enabling accurate identification of both explicit and implicit skill requirements while ensuring comparability across postings.

With skill requirements extracted at the posting level, we construct measures of how firms' skill demands relate to established occupational standards. We assign each job posting to an occupation based on title similarity to O*NET occupational categories, then benchmark the extracted skills against occupation-specific skill sets derived from O*NET task descriptions. Specifically, we map O*NET task descriptions to ESCO skills using semantic similarity, creating a baseline set of "occupation-aligned" skills for each occupation—skills traditionally associated with that occupation under established taxonomies. Skills extracted from job postings that fall within this baseline are classified as occupation-aligned skills; those falling outside are classified as non-aligned skills. This distinction is central to our analysis: occupation-aligned skills capture how precisely firms articulate requirements consistent with existing occupational definitions, while non-aligned skills capture firms' exploration beyond standardized templates.

We measure firm AI capability using the accumulated stock of AI-related patents, constructed via the perpetual inventory method with a 15\% depreciation rate following standard practice in the innovation literature \citep{HallJaffe2000}. AI patents are identified using International Patent Classification (IPC) codes associated with machine learning, neural networks, computer vision, and natural language processing technologies \citep{Yang2022}. This patent-based measure captures firms' underlying technological capabilities in AI, as distinct from short-term hiring intentions or productivity outcomes.

Our baseline empirical specification relates skill outcomes to AI capability at the firm–occupation–year level, controlling for firm, occupation, and year fixed effects. This approach absorbs time-invariant firm characteristics, persistent differences across occupations, and aggregate trends, identifying effects from within-firm variation in AI capability over time. However, a natural concern is that unobserved factors—such as managerial quality or strategic foresight—may simultaneously drive both AI investment and more sophisticated hiring practices. To establish causality, we implement an instrumental variable strategy adapted from \cite{SampatWilliams2019}, exploiting quasi-random variation in patent examiner leniency at China's National Intellectual Property Administration (CNIPA). Patent applications are assigned to examiners through queue-based procedures that are plausibly orthogonal to firm characteristics. We compute each examiner's leniency as their historical grant rate on non-AI patents, then construct a firm-year instrument as the weighted average leniency across examiners assigned to the firm's AI patent applications. Lenient examiners increase the probability that AI patents are granted, generating exogenous variation in firms' AI patent stocks. The exclusion restriction requires that examiner leniency affects skill requirements only through its effect on AI capability, not through direct channels—a condition we probe by examining correlations with non-patent firm outcomes.

Our empirical analysis yields three main findings that speak directly to AI's role in employer signaling. First, we document that AI capability is positively and significantly associated with both occupation-aligned and non-aligned skills listed in job postings. Firms with higher AI stocks specify a richer set of skills that match established occupational standards, and they also list more skills that fall outside these standards. These effects are robust to the inclusion of firm fixed effects and controls, alternative depreciation rates for AI stock construction, and alternative similarity thresholds in the skill mapping procedure. The IV estimates confirm a causal interpretation: exogenous increases in AI capability lead firms to expand their articulated skill portfolios along both dimensions.

Second, we investigate the mechanism through which AI increases occupation-aligned skills, focusing on the hypothesis that AI reduces information asymmetry by helping firms better understand and communicate existing job requirements. If AI improves firms' internal processing of job-related information, we would expect skill descriptions to become more precise and consistent. We test this by examining two measures of textual clarity: (i) text consistency, defined as the within-year similarity of skill descriptions for the same firm–occupation cell, capturing how stably firms describe the same position; and (ii) textual ambiguity, measured by the frequency of vague expressions such as "familiar with," "basic understanding of," or "some knowledge of" in skill-related sentences. Consistent with the information channel, we find that firms with higher AI capability exhibit significantly greater text consistency and significantly lower use of ambiguous language. These patterns suggest that AI helps firms translate internal knowledge into clearer, more structured external signals—reducing noise in the information transmitted to prospective workers.

Third, we examine whether AI enables firms to anticipate evolving skill requirements before they are formally codified in occupational standards. This mechanism reflects a distinct channel: rather than merely clarifying current requirements, AI may help firms identify emerging skill needs that are not yet recognized by official taxonomies. To test this hypothesis, we exploit the fact that occupational classification systems undergo periodic revisions. While our baseline analysis fixes taxonomy definitions to 2018 benchmarks (ESCO v1.0.3 and O*NET v22.3), we can identify skills that were non-aligned under the 2018 taxonomy but became aligned under the 2022 revision (O*NET v27.0). We define these as "forward-looking" skills—requirements that firms demanded before official recognition. If AI enhances firms' ability to anticipate labor market dynamics, AI-capable firms should list more of these forward-looking skills prior to the taxonomy update. Our results strongly support this prediction: firms with higher AI stocks list significantly more forward-looking skills in terms of counts, shares, and intensity. This finding implies that AI allows firms to lead, rather than follow, the formal evolution of occupational standards.

Our results reveal a dual role for AI in employer signaling. On one hand, AI functions as a stabilizer of current recruitment information, helping firms specify existing occupation-specific requirements with greater precision and reducing information asymmetry in the labor market. On the other hand, AI serves as a catalyst for proactive adaptation, enabling firms to detect and respond to emerging skill shifts before they are officially recognized. Both channels contribute to improved signal quality: the first by reducing noise in the communication of established requirements, the second by expanding the informational content of job postings to reflect evolving market realities. These mechanisms stand in contrast to recent findings on the job seeker side, where AI tools that lower the cost of producing signals have been shown to degrade signal informativeness \citep{GaldinSilbert2025, WilesHorton2025}. The key distinction lies in AI's function: in our setting, AI operates as an information technology that improves firms' understanding of their own requirements, rather than merely as a communication technology that reduces the cost of generating text.

This paper contributes to several strands of literature. First, we contribute to the literature on search and matching frictions in labor markets. A foundational body of work establishes that labor markets are characterized by costly search, imperfect information, and endogenous skill requirements \citep{RogersonShimer2005,AlbrechtVroman2002}. Firms strategically adjust skill demands in response to labor supply conditions \cite{ModestinoShoag2016}, and technological change reshapes the skill content of jobs with significant consequences for workers \citep{BraxtonTaska2023}. We contribute by documenting a novel channel through which technology affects labor market efficiency: AI capability improves firms' ability to articulate skill requirements, potentially enhancing match quality by reducing information asymmetry between employers and job seekers. This finding complements recent work showing that AI-assisted communication tools can degrade signaling quality on the job seeker side by making signals cheaper to produce but less informative \citep{GaldinSilbert2025,WilesHorton2025,CuiDias2025}. We show that AI's effect on signaling depends critically on how AI is deployed: when AI functions as an information technology that improves firms' understanding of their own requirements—rather than merely as a writing assistant—it can enhance rather than erode signal quality.

Second, we contribute to the growing literature on AI and labor markets. Existing research has primarily examined AI's effects through the lens of task automation, job displacement, and skill-biased technological change \citep{AcemogluRestrepo2018, AcemogluRestrepo2020, AutorLevy2003, FreyOsborne2017}. More recent work studies AI's implications for firm growth and innovation \citep{Babina2024} and productivity \citep{Yang2022}. We extend this literature by highlighting AI's role as an information technology that shapes how firms process and communicate labor market information. Our findings suggest that AI capability helps firms both clarify current job requirements and anticipate future skill needs—effects that operate through information channels distinct from task automation. To our knowledge, this is among the first studies to document AI's dual function in stabilizing current recruitment signals while catalyzing proactive adaptation to evolving occupational standards.

Third, we make a methodological contribution to the measurement of skill requirements in job vacancy data. Prior work analyzing skills in job postings has relied on keyword matching, dictionary-based approaches, or named entity recognition \citep{DemingKahn2018,HershbeinKahn2018}. These methods struggle with implicit skill expressions and produce measures that are not comparable across postings with different writing styles. Building on recent advances in extreme multi-label classification \citep{Decorte2023,Bhola2020}, we develop a scalable NLP pipeline that maps Chinese job vacancy text to the standardized ESCO skill framework using contrastive learning and LLM-generated synthetic training data. This approach enables consistent measurement of skill demands across firms, occupations, and time, addressing fundamental comparability challenges in labor market text analysis. We make our implementation publicly available to facilitate future research.

Fourth, we provide large-scale firm-level evidence on AI and skill demand from a major emerging economy. The existing literature on technological change and labor markets draws predominantly on data from the United States and Europe. China represents an important setting given its rapid AI adoption, expansive online recruitment ecosystem, and distinct institutional context. Our dataset of 14 million job postings from Chinese listed firms offers granular insight into how AI capability shapes hiring behavior at the firm–occupation level. By combining job vacancy text with patent-based measures of AI capability and employing an instrumental variable strategy based on patent examiner leniency, we establish causal evidence that complements correlational findings from other contexts.

\section{Stylized Facts} \label{sec:Facts}

Our empirical strategy relies on a conceptual distinction between occupation-aligned and non-aligned skills. To motivate this distinction, we first document two stylized facts about occupational classification systems: (1) occupational categories appear stable across major taxonomy revisions, but (2) the task content within occupations changes substantially even when titles remain unchanged. Together, these facts highlight a fundamental tension in labor markets: while the labels used to classify jobs evolve slowly, the actual content of work—and therefore the skills required to perform it—can shift rapidly. This tension raises the central question motivating our analysis: do firms recognize and respond to changes in job content, and does AI capability help them do so?

\subsection{Background: Occupational Classification Systems}
Occupational classification systems are designed and updated under different institutional frameworks, leading to variation in revision frequency, scope, and the manner in which changes are recorded. We examine two major systems (and complement with COCD\footnote{\textbf{COCD} (Chinese Occupational Classification Dictionary) follows a more centralized and periodic revision approach. Since its first release in 1999, COCD has undergone comprehensive revisions in 2015 and 2022, each reflecting broad changes in the Chinese labor market including the emergence of digital and green economy occupations.}) that structure labor market information in different contexts.

\textbf{ESCO} (European Skills, Competences, Qualifications and Occupations) is a multilingual classification maintained by the European Commission. It is continuously updated through versioned releases, including both major revisions—which may introduce new occupations and restructure conceptual categories—and minor updates focused on quality improvements such as label corrections and translation refinements. ESCO provides detailed skill taxonomies that we use as the target framework for our skill extraction.

\textbf{O*NET} (Occupational Information Network) is the primary source of occupational information in the United States, built upon the Standard Occupational Classification (SOC) system. O*NET is updated regularly, with larger structural changes typically occurring when the underlying SOC is revised. Such revisions can involve occupation splits, mergers, and redefinitions, making cross-version comparisons more complex.

While these systems share the goal of describing labor market structures, their revision logic, frequency, and granularity differ considerably. These institutional differences shape how changes manifest over time and provide useful cross-validation for the patterns we document.

\subsection{Stylized Fact 1: Occupational Categories Remain Stable Across Major Revisions}
We first examine whether occupational classifications change substantially across major version updates. Figure 1 compares stability across three systems: ESCO (2018–2022), O*NET (2018–2022), and China's COCD (2015–2022). The plotted percentages represent the share of occupations that remain unchanged across major updates, excluding minor or purely technical edits such as label corrections.

\begin{figure}[!htbp]\centering
\includegraphics[width=0.85\linewidth]{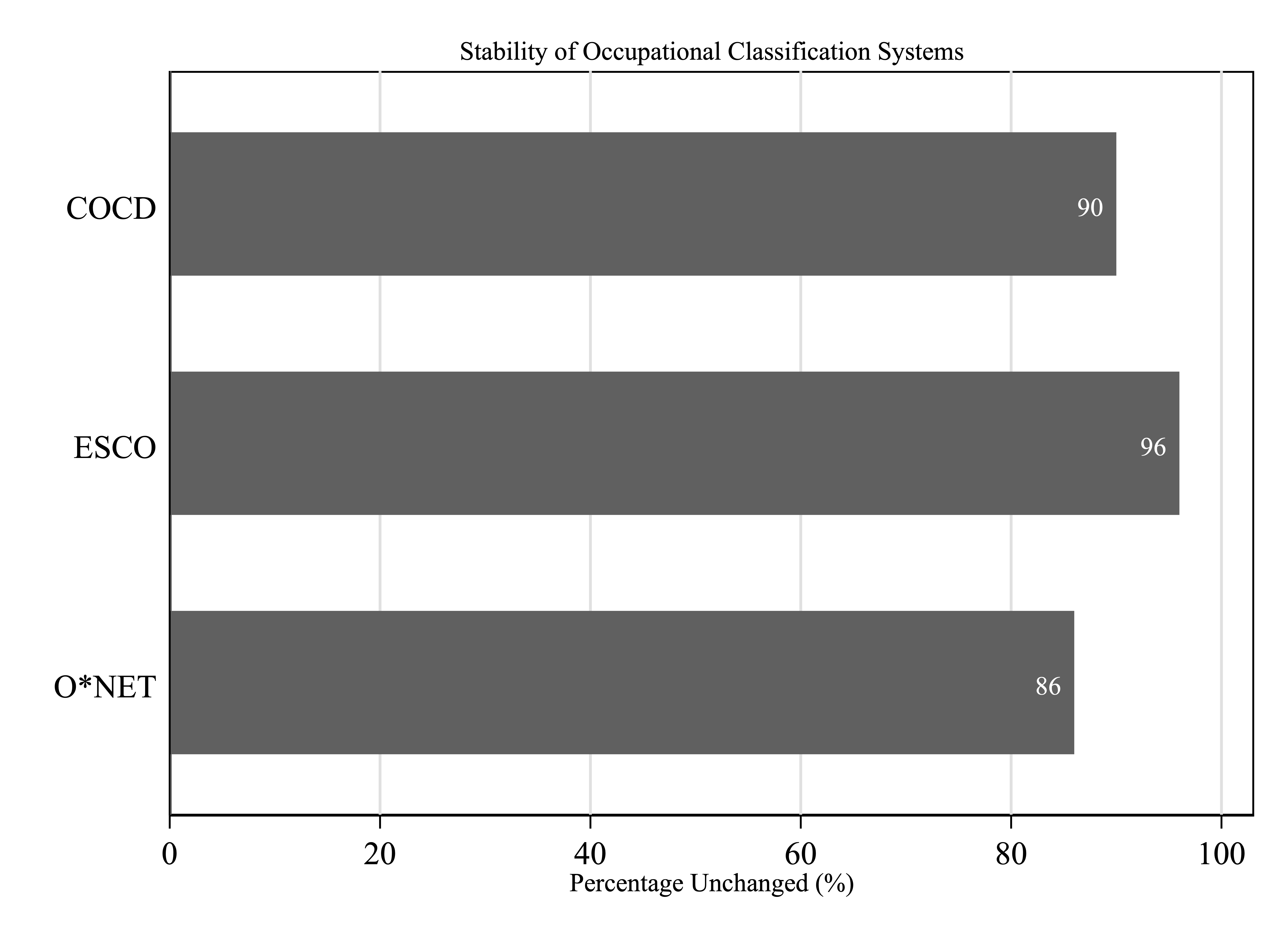}
\caption{Stability across major revisions: ESCO (2018--2022), O*NET (2018--2022), and COCD (2015--2022).}
\label{fig:fig1}
\end{figure}

The evidence shows that occupational lists are remarkably stable even across substantial revisions. ESCO exhibits the highest stability at 96\%, reflecting its continuous maintenance approach where major conceptual changes are relatively rare. COCD shows 90\% stability over a seven-year window that spans significant structural transformation in the Chinese economy. O*NET displays somewhat lower stability at 86\%, partly reflecting SOC-driven restructuring where occupation splits and mergers make measured change more pronounced.

The key takeaway is that occupational taxonomies adjust gradually and through institution-specific pathways. For our empirical design, this stability is valuable: it implies that the set of occupational categories we use to classify job postings does not change dramatically within our analysis window of 2018–2022. Firms posting jobs for "software developers" or "human resources specialists" in 2018 and 2022 are referring to broadly comparable occupational categories.

\subsection{Stylized Fact 2: Task Content Changes Substantially Within Stable Occupations}
While Stylized Fact 1 shows that occupation categories appear stable, this apparent stability may mask substantial changes in job content. To investigate this possibility, we focus on occupations that can be matched one-to-one between O*NET 2018 (version 22.3) and O*NET 2022 (version 27.0), and examine changes in the task sets within each occupation.

We find that approximately 54\% of matched occupations exhibit completely unchanged task sets across this four-year window. The corollary is striking: 46\% of occupations experience at least some task reallocation, even though they retain the same occupational title and code. This task-level dynamism contrasts sharply with the relative stability of occupational classifications documented above.

Figure 2 illustrates occupations with the largest task changes. Task reconfiguration appears across diverse job types. Among routine operational occupations, Packers and Packagers (Hand) and Team Assemblers show substantial task reallocation—likely reflecting automation of certain manual tasks and the addition of new quality control or coordination responsibilities. Among knowledge-intensive professional occupations, Dietitians and Nutritionists and Advertising and Promotions Managers exhibit equally pronounced changes, potentially driven by digitalization, new regulatory requirements, and evolving professional practices.

\begin{figure}[!htbp]\centering
\includegraphics[width=0.70\linewidth]{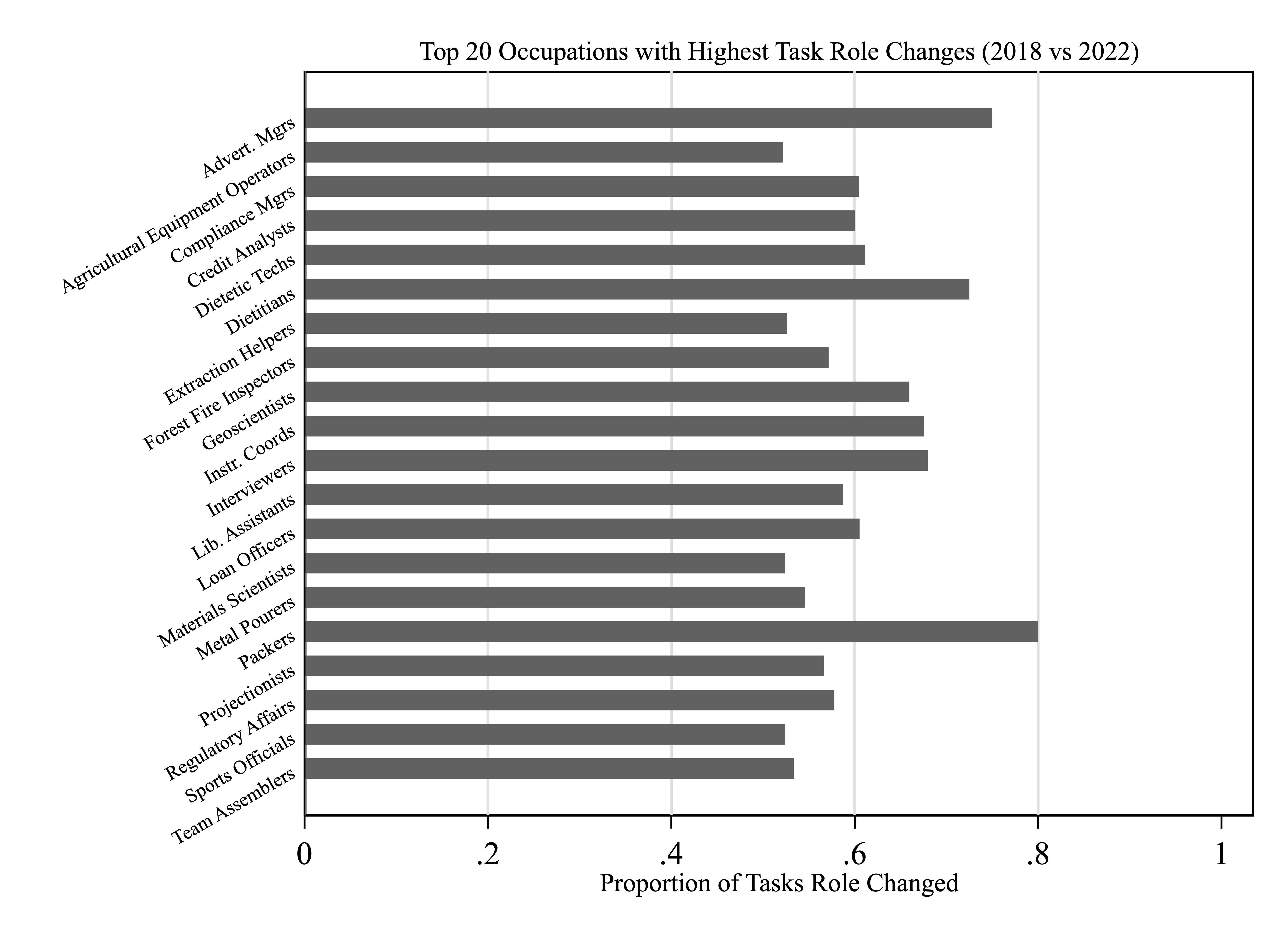}
\caption{Task changes within matched occupations (O*NET 2018--2022): TOP20 occupations with highest task changes.}
\label{fig:fig2}
\end{figure}

In many cases, tasks shift between core and supplementary roles, indicating a rebalancing of work content within stable occupational boundaries. A task that was central to an occupation in 2018 may become peripheral by 2022, while new tasks emerge as primary responsibilities. These patterns suggest that the nature of work is continuously evolving even when the labels we use to describe it remain fixed.

\subsection{Discussion}
These stylized facts carry important implications for understanding employer signaling and skill requirements.
First, the facts highlight that stability in occupation titles should not be interpreted as stability in job content. Labor market change is often reflected more strongly in tasks—and by extension, the skills needed to perform them—than in the occupational labels used to classify jobs. Firms that rely solely on occupational titles to signal their requirements may fail to communicate the actual skills they need.

Second, the facts raise a natural question: do firms recognize and respond to within-occupation changes in skill requirements? If tasks are evolving within occupations, effective employer signaling requires firms to articulate skill demands that go beyond what standard occupational templates would suggest. Firms with superior information-processing capabilities may be better positioned to detect these changes and adjust their job postings accordingly.

Third, the facts motivate our distinction between occupation-aligned and non-aligned skills. By benchmarking job posting requirements against O*NET-derived occupation–skill mappings fixed at the 2018 taxonomy, we can identify skills that conform to established occupational definitions versus skills that fall outside these standards. Given that task content changes substantially between 2018 and 2022, skills classified as non-aligned under the 2018 baseline may include requirements that later become formally recognized—what we term "forward-looking" skills.

Finally, the facts underscore why our analysis focuses on the 2018–2022 window. This period lies between major revision cycles for the taxonomies we use, providing a setting where occupational categories are stable enough to ensure consistent measurement, yet task content is dynamic enough to generate meaningful variation in skill requirements. By fixing taxonomy definitions to the 2018 benchmark, we avoid concerns that observed changes in skill measures are mechanically driven by contemporaneous taxonomy updates.

\section{Data} \label{sec:data}

To investigate the impact of AI technology adoption on skill demand, our analysis leverages three primary datasets: online job vacancy data, patent data, and firm financial data. These datasets enable us to construct detailed measures of skill requirements, AI technology adoption, and firm-level characteristics, focusing on Chinese listed firms and their subsidiaries.

\subsection{Online Job Vacancy Data}
Our job vacancy data are sourced from three leading online recruitment platforms in China: Zhaopin (zhaopin.com), 51job (51job.com), and Liepin (liepin.com). These platforms are among the most widely used job search websites in China, collectively capturing a substantial share of formal sector hiring activity.\footnote{These platforms cater primarily to white-collar and professional occupations. Our sample thus focuses on formal sector employment, which is appropriate for studying listed firms' hiring behavior.} The dataset covers the period from 2015 to 2023 and includes over 200 million job postings, spanning diverse industries, firm sizes, and geographic regions across all 31 provinces in mainland China.

We obtained raw data through a web scraping process and carefully cleaned the dataset to remove duplicate or irrelevant entries. Each posting contains rich unstructured text detailing job titles, responsibilities, and skill requirements, along with structured fields including company names, job locations, posting dates, and salary information. The scale and diversity of this dataset provide a comprehensive representation of labor market skill demands across the Chinese economy, enabling granular analysis at the firm–occupation–year level.

To the best of our knowledge, this dataset represents one of the most comprehensive online recruitment datasets used for academic research in China to date. Its unique features—including large sample size, extensive geographical coverage, and detailed textual content—enable us to conduct rigorous analysis of how firm characteristics shape the skill content of job postings.\footnote{For similar applications of Chinese online job vacancy data, see \citet{Huang2025} and \citet{KuhnShen2012}.}

\subsection{Patent Data}
To measure AI technology adoption, we utilize the Incopat Global Patent Database, which provides comprehensive records of patent applications worldwide. This dataset contains the universe of Chinese patents with detailed information on titles, abstracts, applicants, application dates, and International Patent Classification (IPC) codes.\footnote{Incopat (\url{www.incopat.com}) is widely used in research on Chinese innovation; see \citet{XieZhou2023} for a detailed description.}

Following established approaches in the literature \cite{Yang2022}, we identify AI-related patents based on IPC codes associated with artificial intelligence technologies, including machine learning, neural networks, computer vision, and natural language processing. We use patent applications rather than granted patents because the approval process can take several years, and many AI-related applications filed in recent years remain pending.\footnote{This approach is consistent with \citet{IgnaVenturini2023}. It ensures our focus is on AI-inventing firms rather than AI-using firms, as patent applications reflect real research activity in AI.} Using applicant information from the patent records, we link AI patents to Chinese listed firms and their subsidiaries, enabling us to construct firm-level measures of AI technology capability.

\subsection{ Firm Financial Data}
We incorporate firm financial data from the China Stock Market and Accounting Research (CSMAR) database. CSMAR provides comprehensive information on stock prices, financial statements, corporate governance, and ownership structure for Chinese listed firms. This dataset enables us to control for firm-level characteristics such as size, profitability, leverage, and investment capacity in our empirical analysis.\footnote{CSMAR is the standard source for Chinese listed firm data in academic research; see \citet{YouZhang2022} for a similar application.}

Importantly, CSMAR includes information on the locations of listed firms and their subsidiaries, allowing us to match job vacancy data posted by subsidiary companies to their parent listed firms. This matching process is essential for linking hiring behavior observed in job postings to firm-level AI capability measured through patents.

\section{Methodology} \label{sec:method}

This section describes our approach to measuring skill requirements from job vacancy texts and firm-level AI capability. We first detail the procedure for mapping unstructured job postings to standardized skill and occupation frameworks. We then describe our measure of firm AI capability and present the empirical specification, including an instrumental variable strategy to address endogeneity concerns.

\subsection{Constructing Occupation--Skill Mappings}
\label{subsec:occ_skill_mapping}

A central component of our analysis is the distinction between occupation-aligned and non-aligned skills. To operationalize this distinction, we require a benchmark mapping that defines which skills are traditionally associated with each occupation. We construct this mapping by linking O*NET occupations to ESCO skills through the intermediate layer of task descriptions.

\subsubsection{From O*NET Tasks to ESCO Skills}

O*NET provides detailed task descriptions for each occupation but does not directly enumerate skills. ESCO, by contrast, offers a comprehensive taxonomy of 13,939 skills but does not provide occupation-level linkages comparable to O*NET. We bridge these two frameworks using semantic similarity.

For each occupation $o$ in O*NET, let $\mathcal{T}_o = \{t_1, t_2, \ldots, t_{M_o}\}$ denote the set of associated task descriptions. For each task $t \in \mathcal{T}_o$, we compute its semantic similarity to all ESCO skill descriptions. Specifically, we embed both task descriptions and skill descriptions into a shared vector space using a pre-trained sentence transformer model.\footnote{We use the \texttt{all-MiniLM-L6-v2} model from the Sentence-Transformers library, which maps texts to 384-dimensional dense vectors optimized for semantic similarity tasks.} Let $\mathbf{e}_t \in \mathbb{R}^d$ denote the embedding of task $t$ and $\mathbf{e}_s \in \mathbb{R}^d$ denote the embedding of skill $s$. The similarity between task $t$ and skill $s$ is computed as the cosine similarity:
\begin{equation}
\text{sim}(t, s) = \frac{\mathbf{e}_t^\top \mathbf{e}_s}{\|\mathbf{e}_t\| \cdot \|\mathbf{e}_s\|}.
\end{equation}

For each task $t$, we retain skills whose similarity exceeds a threshold $\tau$:
\begin{equation}
\mathcal{S}_t = \{ s \in \text{ESCO} : \text{sim}(t, s) \geq \tau \}.
\end{equation}
In our baseline specification, we set $\tau = 0.6$ \footnote{Our results are robust to $\tau = 0.5$, $0.7$ and $0.8$.}. The occupation-level baseline skill set is then defined as the union of skills linked to any task within that occupation:
\begin{equation}
\mathcal{B}_o = \bigcup_{t \in \mathcal{T}_o} \mathcal{S}_t.
\end{equation}

This procedure yields, for each O*NET occupation, a set of ESCO skills that represent the competencies traditionally required based on established task structures. We construct baseline mappings using the 2018 versions of both taxonomies (O*NET v22.3 and ESCO v1.0.3) for our main analysis, and separately using the 2022 versions (O*NET v27.0 and ESCO v1.1.1) for mechanism tests involving forward-looking skills.

\subsection{Mapping Job Vacancies to Occupations and Skills}
\label{subsec:vacancy_mapping}

We now describe how we process raw job vacancy texts to extract standardized occupation assignments and skill requirements.

\subsubsection{Step 1: Occupation Assignment}

Each job posting must be assigned to a standardized occupation to enable comparison with baseline skill sets. Job titles in Chinese vacancy data exhibit substantial variation in terminology, abbreviations, and firm-specific naming conventions, making exact matching infeasible.

We address this challenge using approximate nearest neighbor (ANN) search in embedding space. For each job posting $i$, we embed its job title using the same sentence transformer model described above. We similarly embed all O*NET occupation titles. The posting is assigned to the occupation with the highest cosine similarity:
\begin{equation}
o(i) = \arg\max_{o \in \text{O*NET}} \; \text{sim}(\text{title}_i, \text{title}_o).
\end{equation}

To ensure computational efficiency given the scale of our data (over 14 million postings), we implement ANN search using the FAISS library, which enables sub-linear search time through hierarchical indexing.\footnote{FAISS (Facebook AI Similarity Search) implements efficient similarity search for dense vectors. See \citet{JohnsonDouze2021} for technical details.} No similarity threshold is imposed at this stage, as every posting must be assigned to exactly one occupation to construct the firm--occupation--year panel.

\subsubsection{Step 2: Sentence Segmentation}

Job postings typically contain long, loosely structured text blocks describing responsibilities, requirements, and company information. To identify skill requirements at a granular level, we segment each posting into individual sentences.

Standard punctuation-based segmentation performs poorly on Chinese job vacancy texts for several reasons. First, Chinese text does not use spaces between words, and sentence boundaries are often ambiguous. Second, job postings frequently use informal punctuation, including commas, semicolons, and line breaks interchangeably to separate items in lists. Third, skill requirements are often embedded within longer passages describing job responsibilities, requiring semantic rather than purely syntactic parsing.

We therefore employ a BERT-based sentence segmentation model fine-tuned on annotated Chinese corpora. This model treats segmentation as a sequence labeling task, predicting whether each character position represents a sentence boundary based on contextual information. The approach captures semantic coherence rather than relying solely on punctuation marks, yielding more meaningful sentence units for downstream skill extraction.

Let posting $i$ be segmented into a set of sentences:
\begin{equation}
\text{Sentences}_i = \{s_{i1}, s_{i2}, \ldots, s_{iK_i}\},
\end{equation}
where $K_i$ denotes the number of sentences in posting $i$.

\subsubsection{Step 3: Filtering Non-Skill Sentences}

Not all sentences in a job posting describe skill requirements. Many sentences specify educational credentials (e.g., ``Master's degree required''), years of experience, working conditions, or company descriptions. Including such sentences in skill extraction would introduce noise and potentially bias our measures.

We train a binary classification model to identify sentences that contain skill requirements. The classifier is based on a fine-tuned Chinese RoBERTa model (RBT3) that takes a sentence as input and outputs the probability that it describes a skill requirement. Formally, for sentence $s$, the model predicts:
\begin{equation}
p(y = 1 \mid s) = \sigma(\mathbf{w}^\top \mathbf{h}_{\texttt{[CLS]}} + b),
\end{equation}
where $\mathbf{h}_{\texttt{[CLS]}}$ is the hidden state corresponding to the classification token, $\mathbf{w}$ and $b$ are learned parameters, and $\sigma(\cdot)$ is the sigmoid function. Sentences with $p(y=1 \mid s) < 0.5$ are discarded. The training data for this classifier consists of manually labeled sentences from a random sample of job postings, annotated by research assistants with guidelines distinguishing skill descriptions from other content. \footnote{Examples of skill-related sentences include ``Proficient in Python and R for data analysis'' and ``Strong written and verbal communication skills.'' Examples of non-skill sentences include ``Bachelor's degree or above'' and ``Competitive salary and benefits package.''}

\subsubsection{Step 4: Skill Extraction via Extreme Multi-Label Classification}

The core methodological challenge is mapping each skill-related sentence to standardized skills from the ESCO taxonomy. This task presents several difficulties. First, the label space is extremely large: ESCO contains 13,939 distinct skills at the most granular level. Second, skill requirements are often expressed implicitly rather than through exact skill names. For instance, ``ability to independently publish in high-impact international journals'' implies English proficiency and academic writing skills without naming them explicitly \citep{Bhola2020}. Third, a single sentence may reference multiple skills, requiring multi-label prediction.

We address these challenges using an Extreme Multi-Label Classification (XMLC) framework based on contrastive learning, following recent advances in the NLP literature \citep{Liu2017, Decorte2023}. The approach embeds sentences and skill labels into a shared semantic space, enabling retrieval of relevant skills based on embedding similarity.

\paragraph{Overview of the Approach.}
We employ a bi-encoder architecture consisting of two identical encoders: one for job posting sentences and one for ESCO skill descriptions. Each encoder is based on a pre-trained Chinese BERT model augmented with a BiLSTM layer and attention mechanism to capture sequential dependencies. For an input text $x$ (either a sentence or skill description), the encoder produces a normalized embedding $\mathbf{e}_x \in \mathbb{R}^m$ where $\|\mathbf{e}_x\|_2 = 1$. The model is trained using contrastive loss to maximize similarity between matching sentence--skill pairs while minimizing similarity for non-matching pairs.

A key innovation is the use of synthetic training data generated by a large language model (DeepSeek-V3.1). For each ESCO skill, we prompt the LLM to generate realistic job advertisement sentences that would imply that skill at varying proficiency levels (beginner, intermediate, advanced). This augmentation addresses the scarcity of labeled training data and improves the model's ability to recognize implicit skill expressions.

\paragraph{Inference.}
At inference time, for each skill-related sentence $s$ in a job posting, we compute its cosine similarity to all pre-computed ESCO skill embeddings. Skills with similarity exceeding the threshold $\tau = 0.6$ are assigned to the sentence, with each sentence matched to up to five skills to capture multi-skill expressions.

The technical details of the XMLC algorithm---including the bi-encoder architecture, attention mechanism, contrastive loss formulation, and synthetic data generation procedure---are provided in Appendix~\ref{appendix:skill_mapping}. 

\subsubsection{Step 5: Classifying Skills as Aligned or Non-Aligned}

For each job posting $i$ assigned to occupation $o(i)$, we obtain a set of extracted skills $\mathcal{S}_i$ from the XMLC procedure. Each skill $s \in \mathcal{S}_i$ is classified based on whether it belongs to the occupation's baseline skill set:
\begin{itemize}
    \item \textbf{Occupation-aligned skill:} $s \in \mathcal{B}_{o(i)}$
    \item \textbf{Non-aligned skill:} $s \notin \mathcal{B}_{o(i)}$
\end{itemize}

Occupation-aligned skills reflect requirements consistent with established occupational definitions---skills that one would expect based on the occupation's traditional task structure. Non-aligned skills capture requirements that fall outside these baseline definitions, potentially representing firm-specific needs, emerging skill demands, or skills that have not yet been formally codified in occupational standards.

\subsubsection{Step 6: Aggregation to Firm--Occupation--Year Level}

We aggregate posting-level skill measures to the firm--occupation--year level. Let $f$ index firms, $o$ index occupations, and $t$ index years. For each $(f, o, t)$ cell, we compute:
\begin{align}
\text{AlignedSkills}_{f,o,t} &= \sum_{i \in \mathcal{P}_{f,o,t}} |\mathcal{S}_i \cap \mathcal{B}_o|, \\
\text{NonAlignedSkills}_{f,o,t} &= \sum_{i \in \mathcal{P}_{f,o,t}} |\mathcal{S}_i \setminus \mathcal{B}_o|,
\end{align}
where $\mathcal{P}_{f,o,t}$ denotes the set of job postings by firm $f$ for occupation $o$ in year $t$.

These measures capture the total number of aligned and non-aligned skills specified across all postings within each firm--occupation--year cell, forming the basis for our empirical analysis.

\subsection{Measuring Firm AI Capability}
\label{subsec:ai_measure}

We measure firm-level AI capability using patent-based indicators that capture accumulated technological knowledge in artificial intelligence. Compared with alternative measures such as AI-related job postings or survey responses, patent data provide a more direct and stable proxy for firms' underlying technological capabilities.

\subsubsection{Identification of AI Patents}

We identify AI-related patents using the International Patent Classification (IPC) codes commonly adopted in the literature. Following Yang (2022), we classify a patent as AI-related if it contains at least one IPC code associated with artificial intelligence technologies, including machine learning, neural networks, computer vision, and natural language processing. Patent data are obtained from the Incopat global patent database and are matched to Chinese listed firms and their subsidiaries.

\subsubsection{Constructing AI Knowledge Stock}

We measure firm AI capability as the accumulated stock of AI patents using the perpetual inventory method. For firm $f$ in year $t$, AI knowledge stock is constructed as:
\begin{equation}
\text{AI\_stock}_{f,t} = (1 - \delta) \cdot \text{AI\_stock}_{f,t-1} + \text{AI\_flow}_{f,t},
\end{equation}
where $\text{AI\_flow}_{f,t}$ denotes the number of new AI patent applications filed by firm $f$ in year $t$, and $\delta$ is the depreciation rate of AI knowledge.

In our baseline specification, we set $\delta = 0.15$, following standard practice in the literature on intangible capital \citep{HallJaffe2000}. This depreciation rate has been widely used to measure knowledge capital accumulated through R\&D expenditures and is consistent with methodologies employed by the U.S. Bureau of Economic Analysis. In robustness checks, we verify that our results are stable under alternative depreciation rates of $\delta = 0.20$ and $\delta = 0.30$.

\subsection{Empirical Specification}
\label{subsec:empirical_spec}

Our empirical analysis examines how firms' AI capability affects the skill content of job postings at the firm--occupation--year level. The baseline specification is:
\begin{equation}
\ln(\text{Skill}_{f,o,t} + 1) = \beta \cdot \text{AI\_stock}_{f,t} + \gamma_f + \delta_o + \lambda_t + \mathbf{X}_{f,t}'\boldsymbol{\theta} + \varepsilon_{f,o,t},
\label{eq:baseline}
\end{equation}
where $\text{Skill}_{f,o,t}$ denotes the number of skills (aligned or non-aligned) listed by firm $f$ for occupation $o$ in year $t$. We add one before taking logs to accommodate zero values.

The specification includes firm fixed effects ($\gamma_f$), occupation fixed effects ($\delta_o$), and year fixed effects ($\lambda_t$). Firm fixed effects absorb time-invariant firm characteristics such as industry, management quality, and organizational culture. Occupation fixed effects control for persistent differences in skill intensity across job types. Year fixed effects capture aggregate trends in skill reporting and labor market conditions. The vector $\mathbf{X}_{f,t}$ includes time-varying firm-level controls from CSMAR, including firm size (log assets), profitability (ROA), leverage, and R\&D intensity.

The coefficient $\beta$ captures whether within-firm changes in AI capability are associated with changes in skill requirements, holding constant occupation type and year effects. Standard errors are clustered at the firm level to account for serial correlation in hiring behavior.

Our main analysis focuses on the period 2018--2022. This window lies between major revision cycles of occupational taxonomies, limiting concerns that concurrent taxonomy updates mechanically affect our measurement of aligned versus non-aligned skills. By fixing taxonomy definitions to the 2018 benchmark, we ensure that changes in skill classification reflect firm behavior rather than redefinitions of occupational standards.

\subsection{Instrumental Variable Strategy}
\label{subsec:iv}

A natural concern with the baseline specification is that unobserved factors may simultaneously drive AI investment and sophisticated hiring practices. For instance, firms with forward-looking management may both invest more heavily in AI and pay greater attention to articulating skill requirements. To address this endogeneity concern, we implement an instrumental variable strategy based on patent examiner leniency, adapted from \citet{SampatWilliams2019}.

\subsubsection{Conceptual Framework}

The identification strategy exploits quasi-random variation in the probability that a firm's AI patent applications are approved, arising from differences in examiner leniency. Patent applications at China's National Intellectual Property Administration (CNIPA) are assigned to examiners through queue-based procedures that are plausibly orthogonal to the characteristics of individual applications or applicant firms. Examiners vary in their propensity to grant patents due to differences in experience, workload, and interpretation of patentability standards. A firm whose AI applications happen to be assigned to more lenient examiners will, on average, accumulate more granted AI patents---providing exogenous variation in AI capability.

\subsubsection{Construction of the Instrument}

We construct the instrument in two steps. First, we estimate examiner-level leniency using non-AI patents to avoid mechanical correlation with our endogenous variable. For each examiner $e$, leniency is computed as the historical grant rate on non-AI patent applications:
\begin{equation}
Z_e = \frac{1}{N_e} \sum_{p \in \text{non-AI}_e} \mathbf{1}[\text{Granted}_{p,e}],
\end{equation}
where $\text{non-AI}_e$ denotes the set of non-AI patent applications assigned to examiner $e$ during a baseline period (2010--2017), and $\mathbf{1}[\text{Granted}_{p,e}]$ is an indicator for whether application $p$ was granted. Using non-AI patents ensures that examiner leniency is measured independently of the firm's AI patenting activity.

Second, we aggregate examiner leniency to the firm-year level by averaging across all AI patent applications filed by the firm:
\begin{equation}
Z_{f,t} = \frac{1}{|A_{f,t}|} \sum_{a \in A_{f,t}} Z_{e(a)},
\end{equation}
where $A_{f,t}$ denotes the set of AI patent applications filed by firm $f$ in year $t$, and $e(a)$ is the examiner assigned to application $a$. This instrument captures the average leniency of examiners assigned to a firm's AI applications, which affects the probability of patent approval independently of application quality.

\subsubsection{Identification Assumptions}

The validity of this instrument rests on three conditions:

\paragraph{Relevance.} Examiner leniency must predict AI patent grants. Lenient examiners approve a higher share of applications, so firms whose applications are assigned to lenient examiners will accumulate more AI patents. We verify this relationship in first-stage regressions and report F-statistics to assess instrument strength.

\paragraph{Independence.} Examiner assignment must be uncorrelated with firm characteristics that affect skill demand. At CNIPA, patent applications are assigned to examiners based on technological field and arrival order (queue-based assignment), not based on applicant identity or application quality. This institutional feature supports the assumption that examiner assignment is quasi-random conditional on technology class and timing.

\paragraph{Exclusion.} Examiner leniency must affect skill requirements only through its effect on AI patent stock, not through other channels. This assumption would be violated if, for example, lenient examiners were systematically assigned to firms that also happen to have more sophisticated HR practices. We probe this assumption by examining whether examiner leniency predicts firm outcomes unrelated to AI capability.

\section{Results} \label{sec:results}

This section presents our empirical findings. We begin with baseline estimates of the relationship between firm AI capability and skill demand, then assess robustness through alternative specifications and instrumental variable estimation, and finally examine the mechanisms underlying our main results.

\subsection{Baseline Results}
\label{subsec:baseline}

Table~\ref{tab:baseline} reports our baseline regression results examining how firm AI capability affects the skill content of job postings. The dependent variables are the log of one plus the number of occupation-aligned skills (Columns 1--3) and non-aligned skills (Columns 4--6) listed at the firm--occupation--year level. Across specifications, we progressively add firm fixed effects and firm-level controls while always including occupation and year fixed effects.

\begin{table}[!htbp]\centering
\caption{Baseline Results: AI Capability and Skill Demand}
\label{tab:baseline}
\begin{tabular}{lcccccc}
\toprule
 & \multicolumn{3}{c}{\textit{Occupation-aligned skills}} 
 & \multicolumn{3}{c}{\textit{Non-aligned skills}} \\
\cmidrule(lr){2-4}\cmidrule(lr){5-7}
 & (1) & (2) & (3) & (4) & (5) & (6) \\
\midrule
AI\_stock 
& 0.0003*** & 0.0003*** & 0.0002** 
& 0.0003*** & 0.0003*** & 0.0002*** \\
& (0.0001)  & (0.0001)  & (0.0001)
& (0.0001)  & (0.0001)  & (0.0001) \\
\addlinespace
Year FE        & Yes & Yes & Yes & Yes & Yes & Yes \\
Occupation FE  & Yes & Yes & Yes & Yes & Yes & Yes \\
Firm FE        & No  & Yes & Yes & No  & Yes & Yes \\
Controls       & No  & No  & Yes & No  & No  & Yes \\
\midrule
$N$            & 450{,}546 & 450{,}546 & 450{,}538 & 450{,}546 & 450{,}546 & 450{,}538 \\
Adj.\ $R^2$    & 0.446 & 0.446 & 0.506 & 0.206 & 0.207 & 0.311 \\
\bottomrule
\end{tabular}

\vspace{0.2cm}
\begin{minipage}{\linewidth}
\footnotesize
\textit{Notes:} The dependent variables are the log of one plus the number of occupation-aligned skills (Columns 1--3) and non-aligned skills (Columns 4--6) listed in firm job postings at the firm--occupation--year level. AI\_stock is measured using a perpetual inventory method with a 15\% depreciation rate. Controls include log assets, ROA, leverage, and R\&D intensity. Standard errors are clustered at the firm level and reported in parentheses. *** $p<0.01$, ** $p<0.05$, * $p<0.1$.
\end{minipage}
\end{table}

The results reveal a robust positive relationship between AI capability and both types of skill measures. In the most saturated specification (Column 3), a one-unit increase in AI stock is associated with a 0.02\% increase in occupation-aligned skills. While this point estimate may appear modest, the economic magnitude is substantial when evaluated at relevant margins. The standard deviation of AI stock in our sample is approximately 150 patents. A one-standard-deviation increase in AI capability thus corresponds to a $150 \times 0.0002 = 0.03$ log-point increase in aligned skills, or roughly a 3\% increase relative to the mean. For non-aligned skills (Column 6), the same calculation yields a comparable effect size.

To further contextualize these magnitudes, we note that the mean number of aligned skills per firm--occupation--year cell is 12.4, with substantial variation across firms (standard deviation of 8.7). Our estimates imply that moving from the 25th to the 75th percentile of AI capability is associated with approximately 0.4 additional aligned skills per posting---a meaningful increment given that the median posting contains 10 skills. The effect on non-aligned skills is similar in magnitude, suggesting that AI capability expands skill articulation along both dimensions.

The stability of coefficients across specifications is noteworthy. The inclusion of firm fixed effects (Columns 2 and 5) leaves estimates essentially unchanged, indicating that the relationship is driven by within-firm variation over time rather than cross-sectional differences between high-AI and low-AI firms. This pattern alleviates concerns that unobserved firm characteristics simultaneously drive AI investment and sophisticated hiring practices. The addition of time-varying controls (Columns 3 and 6) produces modest attenuation but does not alter the qualitative conclusions.

Taken together, these baseline results indicate that AI capability is associated with both deeper articulation of occupation-aligned skills and broader exploration of skills beyond existing taxonomies. Firms with greater AI capability do not merely list more skills indiscriminately; rather, they specify requirements that are more closely linked to occupational standards while also identifying skill needs that fall outside established definitions. These patterns are consistent with AI functioning as an information technology that improves firms' understanding of job requirements.

\subsection{Robustness and Validation}
\label{subsec:robustness}

We subject our baseline findings to a comprehensive set of robustness checks, including alternative measures of AI capability, alternative similarity thresholds in skill mapping, additional control variables and fixed-effect structures, and instrumental variable estimation.

\subsubsection{Alternative AI Depreciation Rates}

Our baseline measure of firm AI capability is constructed using a perpetual inventory method with a depreciation rate of 15\%. To assess sensitivity to this assumption, we reconstruct AI stock using depreciation rates of 20\% and 30\%, following ranges commonly employed in the literature on intangible capital \citep{AhujaKatila2001, BlundellGriffith1995}.

Table~\ref{tab:robust_ai} reports the results. Across specifications, the estimated coefficients remain positive and statistically significant for both outcome variables. The magnitudes are comparable to baseline estimates, confirming that our findings are not driven by the specific depreciation assumption. Higher depreciation rates mechanically reduce the level of AI stock, but the cross-sectional and temporal variation that identifies our effects is preserved.

\begin{table}[!htbp]\centering
\caption{Robustness: Alternative AI Depreciation Rates}
\label{tab:robust_ai}
\begin{tabular}{lcccc}
\toprule
 & \multicolumn{2}{c}{\textit{Occupation-aligned skills}} 
 & \multicolumn{2}{c}{\textit{Non-aligned skills}} \\
\cmidrule(lr){2-3}\cmidrule(lr){4-5}
 & (1) & (2) & (3) & (4) \\
\midrule
AI stock ($\delta = 0.20$) 
& 0.0002* &  & 0.0003*** &  \\
& (0.0001) &  & (0.0001) &  \\
AI stock ($\delta = 0.30$) 
&  & 0.0002* &  & 0.0003** \\
&  & (0.0001) &  & (0.0001) \\
\addlinespace
Year FE        & Yes & Yes & Yes & Yes \\
Occupation FE  & Yes & Yes & Yes & Yes \\
Firm FE        & Yes & Yes & Yes & Yes \\
Controls       & Yes & Yes & Yes & Yes \\
\midrule
$N$            & 450{,}538 & 450{,}538 & 450{,}538 & 450{,}538 \\
Adj.\ $R^2$    & 0.506 & 0.506 & 0.311 & 0.311 \\
\bottomrule
\end{tabular}

\vspace{0.15cm}
\begin{minipage}{0.95\linewidth}
\footnotesize
\textit{Notes:} The dependent variables are the log of one plus the number of occupation-aligned skills and non-aligned skills. AI stock is constructed using alternative depreciation rates $\delta$. All specifications include firm, occupation, and year fixed effects, plus firm-level controls. Standard errors clustered at the firm level are in parentheses. *** $p<0.01$, ** $p<0.05$, * $p<0.1$.
\end{minipage}
\end{table}

\subsubsection{Alternative Similarity Thresholds}

Our baseline mapping between O*NET task descriptions and ESCO skills relies on a cosine similarity threshold of 0.6. This threshold determines which skills are classified as occupation-aligned versus non-aligned. To ensure results are not artifacts of this specific cutoff, we re-estimate all models using thresholds of 0.5, 0.7, and 0.8.

Table~\ref{tab:robust_similarity} presents the results. Lower thresholds (0.5) produce more permissive mappings with larger baseline skill sets, while higher thresholds (0.7, 0.8) yield more restrictive mappings. Despite these differences in classification, the estimated effects of AI capability remain stable across all thresholds. This robustness reflects the fact that our identification comes from within-firm variation in AI stock, which is orthogonal to the level of the similarity threshold.

\begin{table}[!htbp]\centering
\caption{Robustness: Alternative Similarity Thresholds in Skill Mapping}
\label{tab:robust_similarity}
\begin{tabular}{lcccccc}
\toprule
 & \multicolumn{3}{c}{\textit{Occupation-aligned skills}} 
 & \multicolumn{3}{c}{\textit{Non-aligned skills}} \\
\cmidrule(lr){2-4}\cmidrule(lr){5-7}
 & (1) & (2) & (3) & (4) & (5) & (6) \\
 & $\tau=0.5$ & $\tau=0.7$ & $\tau=0.8$ & $\tau=0.5$ & $\tau=0.7$ & $\tau=0.8$ \\
\midrule
AI\_stock 
& 0.0002** & 0.0002** & 0.0002** 
& 0.0003*** & 0.0003*** & 0.0003*** \\
& (0.0001) & (0.0001) & (0.0001)
& (0.0001) & (0.0001) & (0.0001) \\
\addlinespace
Year FE        & Yes & Yes & Yes & Yes & Yes & Yes \\
Occupation FE  & Yes & Yes & Yes & Yes & Yes & Yes \\
Firm FE        & Yes & Yes & Yes & Yes & Yes & Yes \\
Controls       & Yes & Yes & Yes & Yes & Yes & Yes \\
\midrule
$N$            & 450{,}538 & 450{,}538 & 450{,}538 & 450{,}538 & 450{,}538 & 450{,}538 \\
Adj.\ $R^2$    & 0.615 & 0.488 & 0.484 & 0.325 & 0.309 & 0.312 \\
\bottomrule
\end{tabular}

\vspace{0.15cm}
\begin{minipage}{0.95\linewidth}
\footnotesize
\textit{Notes:} Columns report results using different cosine similarity thresholds $\tau$ for the task--skill mapping that defines occupation-aligned skills. Standard errors clustered at the firm level are in parentheses. *** $p<0.01$, ** $p<0.05$, * $p<0.1$.
\end{minipage}
\end{table}

\subsubsection{Additional Specifications}

We further probe robustness through several additional specifications. First, we control for average posting text length to address the concern that AI-capable firms may simply write longer job descriptions that mechanically contain more skill mentions. Second, we include occupation-by-year fixed effects to absorb occupation-specific trends in skill demand that might correlate with AI adoption patterns. Third, we replace O*NET-based occupation assignments with ESCO occupations to verify that results are not specific to a particular occupational taxonomy.

Table~\ref{tab:robust_other} reports these results. Across all specifications, the positive relationship between AI capability and skill articulation persists. The inclusion of posting length controls leaves estimates essentially unchanged, indicating that effects operate through the \textit{content} rather than merely the \textit{volume} of skill descriptions. Occupation-by-year fixed effects, which absorb any occupation-level shocks to skill demand, produce comparable estimates, suggesting that our findings reflect firm-level responses rather than occupation-level trends.

\begin{table}[!htbp]\centering
\caption{Robustness: Additional Specifications}
\label{tab:robust_other}
\begin{tabular}{lcccccc}
\toprule
 & \multicolumn{3}{c}{\textit{Occupation-aligned skills}} 
 & \multicolumn{3}{c}{\textit{Non-aligned skills}} \\
\cmidrule(lr){2-4}\cmidrule(lr){5-7}
 & (1) & (2) & (3) & (4) & (5) & (6) \\
\midrule
AI\_stock 
& 0.0002** & 0.0002* & 0.0002* 
& 0.0002*** & 0.0002*** & 0.0002*** \\
& (0.0001) & (0.0001) & (0.0001)
& (0.0001) & (0.0001) & (0.0001) \\
\addlinespace
Posting length control & Yes & No & No & Yes & No & No \\
Year FE              & Yes & No & Yes & Yes & No & Yes \\
Occupation FE        & Yes & No & Yes & Yes & No & Yes \\
Occupation $\times$ Year FE & No & Yes & No & No & Yes & No \\
Firm FE              & Yes & Yes & Yes & Yes & Yes & Yes \\
ESCO occupations     & No & No & Yes & No & No & Yes \\
Controls             & Yes & Yes & Yes & Yes & Yes & Yes \\
\midrule
$N$                  & 450{,}538 & 450{,}300 & 450{,}300 & 450{,}538 & 450{,}300 & 450{,}300 \\
Adj.\ $R^2$          & 0.506 & 0.506 & 0.501 & 0.311 & 0.312 & 0.331 \\
\bottomrule
\end{tabular}

\vspace{0.15cm}
\begin{minipage}{0.95\linewidth}
\footnotesize
\textit{Notes:} Column (1) and (4) add controls for average posting text length. Columns (2) and (5) include occupation-by-year fixed effects. Columns (3) and (6) use ESCO-based occupation assignments instead of O*NET. Standard errors clustered at the firm level are in parentheses. *** $p<0.01$, ** $p<0.05$, * $p<0.1$.
\end{minipage}
\end{table}

\subsubsection{Instrumental Variable Estimation}

While the inclusion of firm fixed effects addresses time-invariant confounders, a remaining concern is that time-varying unobservables may simultaneously drive AI investment and hiring sophistication. To establish causality, we implement the instrumental variable strategy described in Section~\ref{subsec:iv}, exploiting quasi-random variation in patent examiner leniency.

Table~\ref{tab:iv_first} reports first-stage results. The instrument---average examiner leniency across a firm's AI patent applications---strongly predicts AI stock. The coefficient is positive and highly significant: firms whose AI applications are assigned to more lenient examiners accumulate larger AI patent portfolios. The first-stage F-statistic exceeds 15, comfortably above conventional thresholds for instrument strength \citep{StockWright2002}. These results confirm that examiner leniency generates meaningful variation in AI capability.

\begin{table}[!htbp]\centering
\caption{Instrumental Variable Estimation: First Stage}
\label{tab:iv_first}
\begin{tabular}{lcc}
\toprule
 & \multicolumn{2}{c}{\textit{Dependent variable: AI\_stock}} \\
\cmidrule(lr){2-3}
 & (1) & (2) \\
\midrule
Examiner leniency ($Z_{f,t}$) 
& 45.832*** & 42.156*** \\
& (11.247) & (10.893) \\
\addlinespace
Year FE        & Yes & Yes \\
Firm FE        & Yes & Yes \\
Controls       & No & Yes \\
\midrule
$N$            & 28{,}428 & 28{,}428 \\
First-stage F  & 16.62 & 14.98 \\
Adj.\ $R^2$    & 0.671 & 0.694 \\
\bottomrule
\end{tabular}

\vspace{0.15cm}
\begin{minipage}{0.85\linewidth}
\footnotesize
\textit{Notes:} The dependent variable is AI\_stock. Examiner leniency is constructed as the weighted average grant rate of examiners assigned to the firm's AI patent applications, where examiner-level leniency is measured using non-AI patents from 2010--2017. Standard errors clustered at the firm level are in parentheses. *** $p<0.01$, ** $p<0.05$, * $p<0.1$.
\end{minipage}
\end{table}

Table~\ref{tab:iv_second} presents the two-stage least squares (2SLS) estimates. Columns (1) and (2) report results for occupation-aligned skills; Columns (3) and (4) report results for non-aligned skills. For comparison, we also present the corresponding OLS estimates.

The IV estimates are positive and statistically significant for both outcome variables. For occupation-aligned skills, the 2SLS coefficient is 0.0003, slightly larger than but statistically indistinguishable from the OLS estimate of 0.0002. For non-aligned skills, the IV estimate is 0.0004, again comparable to the OLS benchmark. The similarity between OLS and IV estimates is reassuring: it suggests that endogeneity bias in the OLS specification is modest, and that our baseline findings can be interpreted causally.

\begin{table}[!htbp]\centering
\caption{Instrumental Variable Estimation: Second Stage (2SLS)}
\label{tab:iv_second}
\begin{tabular}{lcccc}
\toprule
 & \multicolumn{2}{c}{\textit{Occupation-aligned skills}} 
 & \multicolumn{2}{c}{\textit{Non-aligned skills}} \\
\cmidrule(lr){2-3}\cmidrule(lr){4-5}
 & OLS & 2SLS & OLS & 2SLS \\
 & (1) & (2) & (3) & (4) \\
\midrule
AI\_stock 
& 0.0002** & 0.0003** 
& 0.0002*** & 0.0004** \\
& (0.0001) & (0.0001) 
& (0.0001) & (0.0002) \\
\addlinespace
Year FE        & Yes & Yes & Yes & Yes \\
Occupation FE  & Yes & Yes & Yes & Yes \\
Firm FE        & Yes & Yes & Yes & Yes \\
Controls       & Yes & Yes & Yes & Yes \\
\midrule
$N$            & 450{,}538 & 450{,}538 & 450{,}538 & 450{,}538 \\
First-stage F  & --- & 14.98 & --- & 14.98 \\
\bottomrule
\end{tabular}

\vspace{0.15cm}
\begin{minipage}{0.95\linewidth}
\footnotesize
\textit{Notes:} The dependent variables are the log of one plus the number of occupation-aligned skills (Columns 1--2) and non-aligned skills (Columns 3--4). Columns (1) and (3) report OLS estimates; Columns (2) and (4) report 2SLS estimates using examiner leniency as an instrument for AI\_stock. Standard errors clustered at the firm level are in parentheses. *** $p<0.01$, ** $p<0.05$, * $p<0.1$.
\end{minipage}
\end{table}

To probe instrument validity, we conduct several diagnostic tests. First, we examine whether examiner leniency predicts firm outcomes that should be unaffected by AI capability, such as non-AI patent counts or firm size. We find no significant relationship, supporting the exclusion restriction. Second, we verify that examiner assignment is uncorrelated with observable firm characteristics at the time of application, consistent with the quasi-random assignment assumption. Third, we estimate reduced-form regressions of skill outcomes directly on examiner leniency, which yield positive and significant coefficients, confirming that the instrument affects outcomes through the hypothesized channel.\footnote{Detailed results from these diagnostic tests are available upon request.}

Taken together, the IV results support a causal interpretation of our findings: exogenous increases in AI capability lead firms to articulate more skills in job postings, both within and beyond established occupational standards.

\subsection{Mechanisms}
\label{subsec:mechanisms}

Having established the baseline relationship between AI capability and skill articulation, we now examine the mechanisms through which this effect operates. Our theoretical framework suggests two channels: (1) AI reduces information asymmetry, enabling firms to specify current requirements with greater precision; and (2) AI helps firms anticipate evolving skill needs, allowing them to demand emerging skills before official codification. We construct measures to test each mechanism.

\subsubsection{Mechanism 1: Clearer and More Structured Skill Information}

The first mechanism posits that AI capability improves firms' ability to process internal job information and translate it into precise external signals. If this channel is operative, we should observe that AI-capable firms produce more consistent and less ambiguous skill descriptions.

We construct two measures of textual clarity in job postings:

\paragraph{Text Consistency.} For each firm--occupation--year cell, we compute the pairwise cosine similarity among all job postings and take the average. Formally, let $\mathcal{P}_{f,o,t}$ denote the set of postings by firm $f$ for occupation $o$ in year $t$, and let $\mathbf{e}_i$ denote the text embedding of posting $i$. Text consistency is defined as:
\begin{equation}
\text{Consistency}_{f,o,t} = \frac{2}{|\mathcal{P}_{f,o,t}|(|\mathcal{P}_{f,o,t}|-1)} \sum_{i < j} \frac{\mathbf{e}_i^\top \mathbf{e}_j}{\|\mathbf{e}_i\| \|\mathbf{e}_j\|}.
\end{equation}
Higher values indicate that a firm describes the same occupation more uniformly across postings within a year, suggesting a clearer and more stable understanding of job requirements.

\paragraph{Textual Ambiguity.} We measure the prevalence of vague or imprecise language in skill-related sentences. Motivated by the task-based measurement approach in \citet{AutorLevy2003}, we construct text-based indicators of imprecision in skill descriptions. Specifically, we compile a dictionary of ambiguous expressions commonly used in job postings, including phrases such as ``familiar with,'' ``basic understanding of,'' ``some knowledge of,'' ``experience preferred,'' and ``ability to learn.'' For each firm--occupation--year cell, we compute (i) the frequency of ambiguous expressions (count) and (ii) the share of skill sentences containing at least one ambiguous expression. Higher values indicate greater imprecision in skill descriptions.

Table~\ref{tab:mechanism1} reports the results. Column (1) shows that firms with higher AI stock exhibit significantly greater text consistency: a one-standard-deviation increase in AI capability is associated with an 8 percentage point increase in within-cell posting similarity. Columns (2) and (3) show that AI capability is associated with significantly lower textual ambiguity, both in terms of frequency and share. These patterns are consistent with AI improving firms' ability to articulate job requirements clearly, reducing noise and imprecision in the signals transmitted to prospective workers.

\begin{table}[!htbp]\centering
\caption{Mechanism 1: AI Capability and Clarity of Skill Information}
\label{tab:mechanism1}
\begin{tabular}{lccc}
\toprule
 & \multicolumn{3}{c}{\textit{Text clarity measures}} \\
\cmidrule(lr){2-4}
 & (1) & (2) & (3) \\
 & Text consistency 
 & Ambiguity (frequency) 
 & Ambiguity (share) \\
\midrule
AI\_stock 
& 0.0801*** & $-$0.0058*** & $-$0.0025*** \\
& (0.0190)  & (0.0026)   & (0.0011)   \\
\addlinespace
Year FE        & Yes & Yes & Yes \\
Occupation FE  & Yes & Yes & Yes \\
Firm FE        & Yes & Yes & Yes \\
Controls       & Yes & Yes & Yes \\
\midrule
$N$            & 450{,}538 & 450{,}538 & 450{,}538 \\
Adj.\ $R^2$    & 0.518 & 0.371 & 0.313 \\
\bottomrule
\end{tabular}

\vspace{0.15cm}
\begin{minipage}{\linewidth}
\footnotesize
\textit{Notes:} Text consistency measures the average pairwise cosine similarity among job postings within the same firm--occupation--year cell. Ambiguity (frequency) counts vague expressions such as ``familiar with'' and ``basic understanding of'' in skill sentences. Ambiguity (share) is the proportion of skill sentences containing at least one ambiguous expression. Standard errors clustered at the firm level are in parentheses. *** $p<0.01$, ** $p<0.05$, * $p<0.1$.
\end{minipage}
\end{table}

\subsubsection{Mechanism 2: Anticipation of Evolving Skill Requirements}

The second mechanism posits that AI capability helps firms detect and respond to emerging skill needs before they are formally recognized in occupational standards. To test this channel, we exploit the revision of occupational taxonomies between 2018 and 2022.

Recall that our baseline analysis fixes the occupation--skill mapping to the 2018 taxonomy (O*NET v22.3 and ESCO v1.0.3). This allows us to identify skills that were classified as non-aligned under the 2018 benchmark but became aligned under the 2022 revision (O*NET v27.0 and ESCO v1.1.1). We define these as \textit{forward-looking skills}: requirements that firms articulated before official taxonomies recognized them as occupation-relevant.

Formally, for occupation $o$, let $\mathcal{B}_o^{2018}$ denote the baseline skill set under the 2018 taxonomy and $\mathcal{B}_o^{2022}$ denote the baseline skill set under the 2022 taxonomy. Forward-looking skills are defined as:
\begin{equation}
\mathcal{F}_o = \mathcal{B}_o^{2022} \setminus \mathcal{B}_o^{2018},
\end{equation}
i.e., skills that are linked to occupation $o$ under the 2022 taxonomy but were not linked under the 2018 taxonomy. These skills reflect task and competency redefinitions that were only formally codified in later taxonomy updates.

If AI enables firms to anticipate labor market dynamics, firms with higher AI capability should list more forward-looking skills \textit{before} the 2022 taxonomy revision. We test this hypothesis using three outcome measures:

\begin{itemize}
    \item \textit{Forward-looking skill count}: The number of forward-looking skills appearing in a firm's job postings for occupation $o$ in year $t$, computed as $|\mathcal{S}_{f,o,t} \cap \mathcal{F}_o|$.
    \item \textit{Forward-looking skill share}: The ratio of forward-looking skills to total non-aligned skills (under the 2018 taxonomy), measuring the composition of non-aligned skill demand.
    \item \textit{Forward-looking skill intensity}: The average number of forward-looking skill mentions per posting within the firm--occupation--year cell.
\end{itemize}

Table~\ref{tab:mechanism2} presents the results. Across all three measures, firms with higher AI capability list significantly more forward-looking skills. Column (1) shows that a one-standard-deviation increase in AI stock is associated with approximately 1.2 additional forward-looking skills per firm--occupation--year cell. Columns (2) and (3) confirm that this effect operates through both the extensive margin (more forward-looking skills overall) and the intensive margin (more mentions per posting).

\begin{table}[!htbp]\centering
\caption{Mechanism 2: AI Capability and Forward-Looking Skills}
\label{tab:mechanism2}
\begin{tabular}{lccc}
\toprule
 & \multicolumn{3}{c}{\textit{Forward-looking skill measures}} \\
\cmidrule(lr){2-4}
 & (1) & (2) & (3) \\
 & Count 
 & Share 
 & Intensity \\
\midrule
AI\_stock 
& 0.0080*** & 0.0085*** & 0.0085*** \\
& (0.0021)  & (0.0029)  & (0.0025)  \\
\addlinespace
Year FE        & Yes & Yes & Yes \\
Occupation FE  & Yes & Yes & Yes \\
Firm FE        & Yes & Yes & Yes \\
Controls       & Yes & Yes & Yes \\
\midrule
$N$            & 450{,}538 & 450{,}538 & 450{,}538 \\
Adj.\ $R^2$    & 0.766 & 0.766 & 0.766 \\
\bottomrule
\end{tabular}

\vspace{0.15cm}
\begin{minipage}{\linewidth}
\footnotesize
\textit{Notes:} Forward-looking skills are defined as skills classified as non-aligned under the 2018 taxonomy (O*NET v22.3, ESCO v1.0.3) but aligned under the 2022 taxonomy (O*NET v27.0, ESCO v1.1.1). Count is the number of forward-looking skills in the firm--occupation--year cell. Share is the ratio of forward-looking skills to total non-aligned skills. Intensity is the average number of forward-looking skill mentions per posting. Standard errors clustered at the firm level are in parentheses. *** $p<0.01$, ** $p<0.05$, * $p<0.1$.
\end{minipage}
\end{table}

These findings suggest that AI capability enables firms to identify and articulate emerging skill requirements ahead of their formal codification in occupational standards. Rather than passively following taxonomy updates, AI-capable firms appear to \textit{lead} the evolution of skill definitions---detecting shifts in job content and adjusting their hiring signals accordingly. This pattern is consistent with AI functioning as a tool for processing labor market information and anticipating future trends, complementing its role in clarifying current requirements documented in Mechanism 1.

\section{Discussion} \label{sec:discussion}

We have presented three central findings in this paper. First, firm AI capability is positively and significantly associated with the articulation of both occupation-aligned and non-aligned skills in job postings. This relationship is robust across alternative measures of AI capability, similarity thresholds in skill mapping, and fixed-effect structures, and is confirmed by instrumental variable estimation exploiting quasi-random variation in patent examiner leniency. Second, AI capability improves the clarity and consistency of skill descriptions: firms with greater AI stocks produce more uniform job postings for the same occupation and use less ambiguous language when specifying requirements. Third, AI-capable firms list more forward-looking skills---requirements that are not yet recognized by official occupational taxonomies but will be codified in future revisions---suggesting that AI enables firms to anticipate evolving labor market dynamics rather than merely responding to established standards.

This robust positive link between AI capability and skill articulation may arise through multiple channels. Our mechanism analyses provide evidence for two distinct pathways. The clarity channel suggests that AI functions as an information-processing technology that helps firms translate internal organizational knowledge into precise external signals. Firms possess tacit understanding of the competencies required for their positions, but articulating this knowledge in job postings requires cognitive effort and may be subject to noise and inconsistency. AI tools---including natural language processing, knowledge management systems, and data analytics---may reduce these frictions by helping HR departments codify job requirements more systematically. The evidence that AI-capable firms produce more consistent postings and use less vague language supports this interpretation.

The anticipation channel suggests a more forward-looking role for AI in labor market signaling. Occupational classification systems, while valuable for structuring labor market information, are inherently backward-looking: they codify skill requirements based on established practices rather than emerging needs. The stylized facts documented in Section~\ref{sec:Facts} illustrate this tension---occupational titles remain stable while task content evolves substantially. AI capability may help firms detect these within-occupation changes earlier, either by processing signals from their own operations (e.g., identifying which skills predict employee performance) or by analyzing external information (e.g., monitoring technological trends and competitor practices). Our finding that AI-capable firms list more forward-looking skills---requirements that will later be recognized by official taxonomies---is consistent with this interpretation.

These findings contribute to ongoing debates about AI's role in labor markets. A prominent strand of recent research documents that AI-assisted communication tools can degrade signaling quality by making signals cheaper to produce but less informative \citep{GaldinSilbert2025, CuiDias2025, WilesHorton2025}. In these settings, AI functions primarily as a \textit{writing technology} that reduces the cost of generating text without necessarily improving the underlying information being communicated. Our findings suggest a complementary perspective: when AI functions as an \textit{information technology} that improves firms' understanding of their own requirements, it can enhance rather than erode signal quality. The key distinction lies not in whether AI is used, but in \textit{how} AI capability is deployed---for superficial communication assistance versus substantive information processing.

This distinction carries implications for understanding AI's asymmetric effects on the two sides of the labor market. On the job seeker side, AI writing tools lower the cost of producing application materials without necessarily improving applicants' underlying qualifications or fit. This creates a classic signal-jamming problem: when everyone can produce polished applications at low cost, the informativeness of application quality as a signal of ability declines \citep{Spence1973}. On the employer side, by contrast, AI capability appears to operate through a different mechanism---improving firms' actual understanding of job requirements rather than merely their ability to describe them. This asymmetry suggests that AI's net effect on labor market matching efficiency depends on the relative magnitudes of signal degradation on the worker side and signal enhancement on the employer side, an empirical question that merits further investigation.

Our findings also speak to the literature on technological change and skill requirements. Prior work has documented secular trends in skill demand, including the rising importance of cognitive and social skills \citep{AutorLevy2003, Deming2017} and the emergence of new skill categories associated with digitalization \citep{Acemoglu2022}. These studies typically analyze skill trends at the occupation or industry level, treating skill requirements as shaped by technological characteristics of jobs. Our contribution is to highlight the role of firm-level technological capability in shaping how skill requirements are \textit{communicated}. Even holding occupation constant, firms with greater AI capability articulate different---and arguably better---skill signals. This suggests that observed trends in skill demand may partly reflect improvements in firms' ability to express their requirements, not only changes in the underlying nature of work.

The mechanism evidence on forward-looking skills has implications for understanding the co-evolution of firm practices and occupational standards. Official taxonomies such as O*NET and ESCO are constructed through systematic surveys, expert panels, and stakeholder consultation---processes that inherently lag behind real-time changes in job content \citep{AutorDorn2013}. Our finding that AI-capable firms anticipate taxonomy revisions suggests that firm hiring behavior may contain leading indicators of occupational change. Policymakers responsible for maintaining occupational standards might leverage such signals---for instance, by monitoring skill demands in job postings from technologically advanced firms---to accelerate taxonomy updates and reduce lags in labor market information infrastructure.

At the same time, our findings raise questions about potential distributional consequences. If AI capability enables some firms to signal more effectively while others continue to rely on imprecise or outdated job descriptions, this heterogeneity could affect matching outcomes. Workers may sort more efficiently toward AI-capable firms that clearly articulate their requirements, while firms lacking AI capability may experience worse matches and higher turnover. To the extent that AI capability correlates with firm size, industry, or geography, these dynamics could exacerbate existing inequalities in labor market access and match quality. Further research is needed to trace these downstream consequences.

\subsection{Limitations}

Several limitations of our analysis merit discussion. First, our sample is restricted to Chinese listed firms and their subsidiaries, which represent a selected segment of the labor market. Listed firms tend to be larger, more formally organized, and more likely to invest in advanced technologies than privately held companies or small businesses. While this sample selection is necessary to construct firm-level measures of AI capability from patent data, it limits the generalizability of our findings to the broader population of employers. The relationship between AI capability and skill articulation may differ for smaller firms, informal sector employers, or firms in different institutional contexts.

Second, our measure of AI capability based on patent stocks captures firms' \textit{inventive} activity in AI but may not fully reflect their \textit{use} of AI tools in HR and recruitment processes. A firm with many AI patents focused on computer vision, for instance, may not apply AI to its hiring practices. Conversely, firms that purchase and deploy AI-powered HR software without developing proprietary AI technology would not be captured by our patent-based measure. This measurement limitation likely attenuates our estimates, as some firms classified as having low AI capability may nonetheless use AI in recruitment. Alternative measures---such as AI-related job postings in HR functions or survey data on AI adoption---could complement our approach in future research.

Third, while our instrumental variable strategy addresses endogeneity from unobserved firm characteristics, it identifies a local average treatment effect for firms whose AI patenting is affected by examiner leniency. This complier population may not be representative of all firms, and the estimated effects may not generalize to interventions that affect AI capability through different channels. Additionally, the exclusion restriction---that examiner leniency affects skill requirements only through AI stock---cannot be directly tested and relies on institutional assumptions about the patent examination process.

Fourth, our analysis focuses on the \textit{quantity} and \textit{precision} of skill signals but does not directly observe labor market outcomes such as application quality, hiring success, or match duration. While clearer skill signals should theoretically improve matching efficiency, this conjecture awaits direct empirical validation. Future work linking job posting characteristics to hiring outcomes would provide more definitive evidence on the welfare implications of AI-enhanced employer signaling.

Finally, our data predate the widespread deployment of large language models (LLMs) such as ChatGPT, which became publicly available in late 2022. The AI capabilities captured by our patent-based measure reflect earlier generations of AI technology, including machine learning, computer vision, and natural language processing systems developed through substantial R\&D investment. The effects of more recent generative AI tools---which dramatically lower the cost of producing text and may be adopted rapidly even by firms without prior AI capability---could differ substantially from the patterns we document. Whether LLM-assisted job posting creation enhances or degrades signal quality is an important question for future research.

\section{Conclusion}
\label{sec:conclusion}

This paper examines how firm-level AI capability affects employer signaling in labor markets, focusing on the skill content of job postings. Using 14 million online job vacancies from Chinese listed firms matched to AI patent data, we develop a novel methodology for extracting and classifying skills based on extreme multi-label classification with contrastive learning. Our empirical analysis documents a robust causal relationship between AI capability and expanded skill articulation, operating through two distinct mechanisms: clearer specification of current requirements and earlier adoption of forward-looking skills.

These findings highlight AI's potential to function as an information technology that improves labor market efficiency, complementing recent evidence on AI's role in degrading signals on the job seeker side. The key insight is that AI's effect on signaling quality depends on how it is deployed: AI that reduces the cost of producing signals without improving underlying information may erode informativeness, while AI that enhances firms' understanding of their own requirements can strengthen the quality of labor market communication.

Our results carry implications for both research and policy. For researchers, we demonstrate the value of analyzing employer signaling through the lens of skill articulation, moving beyond aggregate measures of labor demand to examine how firms communicate their requirements. The XMLC methodology we develop provides a scalable approach for extracting comparable skill measures from large-scale text data, with potential applications in other contexts where unstructured job postings must be mapped to standardized frameworks. For policymakers, our findings suggest that promoting AI adoption among employers could yield benefits for labor market matching, though attention to distributional consequences is warranted if AI capability remains concentrated among larger or more technologically advanced firms.

As AI continues to transform both the demand for and communication of skills in labor markets, understanding these dynamics becomes increasingly important. Our paper provides an initial framework for analyzing AI's role in employer signaling, while highlighting the need for continued research on how technological change shapes the information environment in which workers and firms find each other.

\newpage{}

\bibliographystyle{aer}
\bibliography{ref}

\newpage{}

\appendix

\counterwithin{figure}{section}
\counterwithin{table}{section}

\begin{appendices}
\pagenumbering{arabic}%

{\centerline{\huge{\textbf{Appendices for Online Publication Only}}}}

\bigskip

{\centerline{\large{Hangyu Chen, Yongming Sun, Yiming Yuan}}}

\bigskip

{\centerline{\large{January, 2026}}}

% %%appendix a

%\renewcommand\thefigure{\thesection.\arabic{figure}}  

\section{Mapping Real World Job Vacancies to ESCO Skills}\label{appendix:skill_mapping}
In this section, we delineate the overarching framework of our project, which leverages an eXtreme Multi-Label Classification (XMLC) structure to identify skills from job descriptions. The primary objective is to map textual sentences to a vast set of potential skills, drawn from the ESCO (European Skills, Competences, Qualifications and Occupations) taxonomy. Our approach integrates synthetic data generation via Large Language Models (LLMs), a pre-screening binary classification step, and a bi-encoder model trained with contrastive loss for precise skill matching. We emphasize the methodological and mathematical underpinnings, highlighting innovations in data augmentation, filtering, and model training.

\subsection{General Framework}

The framework comprises three core stages: data preparation, pre-screening, and skill identification via XMLC.

\textbf{Data Preparation:} We begin by generating synthetic training data to augment the limited labeled corpus. Utilizing an LLM (specifically, DeepSeek-V3.1), we create diverse sentences that simulate job advertisement contexts for each ESCO skill at varying proficiency levels (beginner, intermediate, advanced). This process ensures a rich, balanced dataset that captures nuanced skill requirements without relying solely on real-world data, which may suffer from sparsity or bias.

\textbf{Pre-Screening:} To enhance efficiency and accuracy in XMLC, we employ a BERT-based binary classifier to filter sentences. This step distinguishes sentences containing skill requirements from those that do not, reducing noise in downstream processing. The classifier is fine-tuned on a labeled dataset, enabling sentence-level predictions.

\textbf{Skill Identification via XMLC:} The core component is a bi-encoder architecture trained for multi-label classification. Sentences are encoded into embeddings and matched against skill embeddings from the ESCO taxonomy. This setup handles the extreme scale of labels (thousands of skills) by computing similarities in embedding space, facilitating top-k retrieval for skill recommendations.

The framework processes job descriptions by first segmenting them into sentences using a BERT-based tokenizer, applying binary classification to select relevant sentences, and then performing XMLC to assign skills.

\subsection{Mathematical Model}

Our model is grounded in embedding-based retrieval and contrastive learning, formalized as follows.

\subsubsection{Bi-Encoder Architecture}

The bi-encoder consists of two identical encoders: one for sentences and one for skills. Let $\mathcal{E}$ denote the encoder function, parameterized by a BERT backbone with additional layers. For a sentence $s$ and a skill description $d_k$ (from ESCO), their embeddings are:
\[
\mathbf{e}_s = \mathcal{E}(s), \quad \mathbf{e}_k = \mathcal{E}(d_k),
\]
where $\mathbf{e}_s, \mathbf{e}_k \in \mathbb{R}^m$ are normalized vectors ($\|\mathbf{e}\|_2 = 1$), and $m$ is the embedding dimension (e.g., 128).

To capture sequential dependencies, we incorporate a BiLSTM layer followed by an attention mechanism. The BERT output yields hidden states $\mathbf{H} = \{\mathbf{h}_1, \dots, \mathbf{h}_L\} \in \mathbb{R}^{L \times h}$, where $L$ is the sequence length and $h$ is the hidden size. The BiLSTM processes $\mathbf{H}$:
\[
\mathbf{O} = \text{BiLSTM}(\mathbf{H}) \in \mathbb{R}^{L \times 2b},
\]
with $b$ as the LSTM hidden size. Attention weights $\boldsymbol{\alpha} \in \mathbb{R}^L$ are computed as:
\[
\mathbf{u}_t = \tanh(\mathbf{W}_a \mathbf{o}_t + \mathbf{b}_a), \quad \alpha_t = \frac{\exp(\mathbf{v}_a^\top \mathbf{u}_t)}{\sum_{j=1}^L \exp(\mathbf{v}_a^\top \mathbf{u}_j)},
\]
where $\mathbf{W}_a \in \mathbb{R}^{a \times 2b}$, $\mathbf{b}_a \in \mathbb{R}^a$, $\mathbf{v}_a \in \mathbb{R}^a$, and $a$ is the attention dimension. The context vector is $\mathbf{c} = \sum_{t=1}^L \alpha_t \mathbf{o}_t$, projected to $\mathbf{e}$ via a linear layer.

\subsubsection{Contrastive Loss}

Training employs contrastive loss to align embeddings of matching sentence-skill pairs while repelling non-matches. For a positive pair $(\mathbf{e}_s, \mathbf{e}_k^+)$ and negative pairs $\{(\mathbf{e}_s, \mathbf{e}_k^-)_j\}_{j=1}^N$, the loss is:
\[
\mathcal{L} = \frac{1}{B} \sum_{i=1}^B \left[ (1 - \mathbf{e}_{s_i}^\top \mathbf{e}_{k_i^+}) + \frac{1}{N} \sum_{j=1}^N \max(0, \mathbf{e}_{s_i}^\top \mathbf{e}_{k_{i,j}^-} - \gamma) \right],
\]
where $B$ is the batch size, $N$ is the number of negatives, and $\gamma > 0$ is the margin (e.g., 0.5). This encourages $\mathbf{e}_s^\top \mathbf{e}_k^+ \approx 1$ and $\mathbf{e}_s^\top \mathbf{e}_k^- < \gamma$.

For multi-label cases (sentences with multiple skills), the loss is averaged over all positive labels per sample.

\subsubsection{Inference and Ranking}

At inference, for a sentence $s$, compute similarities $\sigma_k = \mathbf{e}_s^\top \mathbf{e}_k$ for all skills $k \in \{1, \dots, K\}$. Rank skills by descending $\sigma_k$ and select top-$r$ (e.g., $r=5$). Evaluation uses Mean Reciprocal Rank (MRR):
\[
\text{MRR} = \frac{1}{Q} \sum_{q=1}^Q \frac{1}{\text{rank}_q},
\]
and Recall@5 (R@5), where $Q$ is the number of queries.

\subsection{Improvements and Enhancements}

We introduce several enhancements to address challenges in skill identification, such as data scarcity, noise, and multi-label complexity.

\textbf{Synthetic Data Generation via LLM:} To mitigate label sparsity in XMLC, we prompt an LLM to generate sentences for each ESCO skill at three proficiency levels. The prompt includes skill definitions and level-specific guidelines, yielding diverse, contextually rich examples. This augmentation expands the training set, improving model robustness without manual labeling.

\textbf{Binary Pre-Screening Classifier:} XMLC struggles with non-skill sentences, leading to false positives. We fine-tune a RoBERTa-base (rbt3) model for binary classification: label 1 if a sentence contains skill requirements, 0 otherwise. The model is trained with cross-entropy loss:
\[
\mathcal{L}_\text{bin} = -\frac{1}{B} \sum_{i=1}^B \left[ y_i \log(p_i) + (1-y_i) \log(1-p_i) \right],
\]
where $y_i \in \{0,1\}$ and $p_i$ is the predicted probability. This filters irrelevant sentences, focusing XMLC on high-potential inputs and boosting overall precision.

\textbf{Sentence-Level Processing with BERT Segmentation:} Job descriptions are segmented into sentences using a BERT tokenizer, enabling granular predictions. This approach captures localized skill mentions, improving accuracy over document-level methods.

\textbf{Multi-Label Handling in Bi-Encoder:} We extend the model to multi-label scenarios by treating each sentence as associated with multiple positive skills during training. The contrastive loss is computed per positive label, averaging contributions, which enhances the model's ability to discern overlapping skills.

\textbf{Negative Sampling Strategy:} Hard negatives are sampled randomly from non-matching skills, with multiple negatives per positive to enrich the loss signal. This promotes better separation in embedding space.

These improvements collectively yield a more accurate and efficient XMLC system for skill extraction, as validated by enhanced MRR and R@5 metrics.
\setcounter{figure}{0}  

% \noindent T

\end{appendices}

\end{document}